\documentclass[aps,pra,twocolumn,showpacs]{revtex4}
\usepackage{epsfig}
\usepackage{amsbsy,latexsym}
\usepackage{amsmath}
\usepackage{amssymb, mathrsfs}
\usepackage[mathscr]{eucal}
\usepackage{hyperref}

\def\d{\operatorname{d}}\def\<{\langle}\def\>{\rangle}
\def\Tr{\operatorname{Tr}}\def\:{\hbox{\bf :}}

\def\set#1{{\sf #1}}
\def\dag{\dagger}
\def\geq{\geqslant}
\def\map#1{\mathcal #1}
\def\sH{\mathcal{H}}\def\sK{\mathcal{K}}

\def\qed{$\,\blacksquare$\par}
\def\kk{\rangle\!\rangle}\def\bb{\langle\!\langle}

\newtheorem{lemma}{Lemma}

\newtheorem{theorem}{Theorem}

\newcommand{\ket}[1]{| #1 \rangle}
\newcommand{\bra}[1]{\langle #1 |}
\newcommand{\Ket}[1]{| #1 \rangle \! \rangle}
\newcommand{\Bra}[1]{\langle \! \langle #1 |}

\newcommand{\hilb}[1]{\mathcal{#1}}

\begin{document}
\title{Optimal Quantum Learning of a Unitary Transformation}

\author{Alessandro Bisio} 
\affiliation{QUIT Group,  Dipartimento di Fisica ``A. Volta'' and INFN Sezione di Pavia, via Bassi
  6, 27100 Pavia, Italy.} 

\author{Giulio Chiribella} 
\affiliation{Perimeter Institute for Theoretical Physics, 31 Caroline Street North, Waterloo, Ontario N2L 2Y5, Canada.} 

\author{Giacomo Mauro D'Ariano}
\affiliation{QUIT Group,  Dipartimento di Fisica ``A. Volta'', INFN Sezione di Pavia, via Bassi
  6, 27100 Pavia, Italy.} 
  
\author{Stefano Facchini} 
\affiliation{QUIT Group,  Dipartimento di Fisica ``A. Volta'', INFN Sezione di Pavia, via Bassi
  6, 27100 Pavia, Italy.} 
  
\author{Paolo Perinotti} 
\affiliation{QUIT Group,  Dipartimento di Fisica ``A. Volta'', INFN Sezione di Pavia, via Bassi
  6, 27100 Pavia, Italy.} 
\date{ \today}
\begin{abstract} 
  We address the problem of learning an unknown unitary transformation
  from a finite number of examples. The problem consists in finding the learning machine that optimally emulates the examples, thus reproducing the unknown unitary maximum
  fidelity. Learning a unitary is equivalent to storing it in the
  state of a quantum memory (the memory of the learning machine), and
  subsequently retrieving it.  We prove that, whenever the unknown
  unitary is drawn from a group, the optimal strategy consists in a
  parallel call of the available uses followed by a ``measure-and-rotate'' retrieving.
  Differing from the case of quantum cloning, where the incoherent
  ``measure-and-prepare'' strategies are typically suboptimal, in the
  case of learning the ``measure-and-rotate" strategy is optimal even when
  the learning machine is asked to reproduce a single copy of the
  unknown unitary.  We finally address the problem of the optimal
  inversion of an unknown unitary evolution, showing also in this
  case the optimality of the ``measure-and-rotate'' strategies and
  applying our result to the optimal approximate realignment of
  reference frames for quantum communication.
  \end{abstract}
\maketitle

\section{Introduction}

A quantum memory would be an invaluable resource for Quantum
Technology, and extensive experimental work is in progress for its
realization \cite{qmem1,qmem2,qmem3}. On a quantum memory one can
store unknown quantum states. Can we exploit it to store an unknown
quantum transformation? In this way we could transmit the
transformation to a distant party by just transmitting a state,
without the need of transferring the device. More generally, we could
process the transformation with the usual state manipulation
techniques, as noticed by Vidal, Masanes, and Cirac, who addressed the problem in Ref.  \cite{cirac}.

Storing-retrieving of transformations can also be seen as an instance
of \emph{quantum learning}, a topic which received increasing
attention in the past few years (see e.g. Refs.
\cite{sasaki,sasacarl,moelmer} for different approaches):
Suppose that a user can dispose of $N$ uses of a black box
implementing an unknown unitary transformation $U$.  Today the user
is allowed to exploit the black box at his convenience, running an
arbitrary quantum circuit that makes $N$ calls to it.  Tomorrow,
however, the black box will no longer be available, and the user
will be asked to reproduce $U$ on a new input state $|\psi\>$ unknown
to him of her. We refer to this scenario as to quantum learning of the
unitary $U$ from a finite set of $N$ examples.  Generally, the user
may be required to reproduce $U$ more than once,  i.e. to produce
$M\ge 1$ copies of $U$. In this case it is important to assess how the
performance of learning decays with the number of copies required, as it was done in the case
of quantum cloning \cite{clonrev}.

Let us consider first the $M=N=1$ case.  Clearly, the only thing  we can do today is to apply the black box to a known
(generally entangled) state $|\varphi\>$.  After that, what
remains is the state $|\varphi_U\>=(U\otimes
I)|\varphi\>$, that can be stored in a quantum memory.  Then, when the new input
state $|\psi\>$ becomes available, we send $|\psi\>$ and
$|\varphi_U\>$ to an optimal \emph{retrieving channel}, which emulates
$U$ applied to $|\psi\>$.  If $N >1$ input copies are available, we must also find the best storing strategy: we can, e.~g.,  opt for a
{\em parallel strategy} where $U$ is applied on $N$ different systems,
yielding $(U^{\otimes N } \otimes I)|\varphi\>$, or for a {\em
  sequential strategy} where $U$ is applied $N$ times on the same
system, alternated with other known unitaries, yielding $(U
V_{N-1}\ldots V_2UV_1U \otimes I) |\varphi\>$.  The most general
storing strategy is described by a \emph{quantum circuit board}, {\em
  i.e.},  a quantum network with open slots where the input copies
can be inserted \cite{architecture,comblong}.
 In summary, solving the
problem of the optimal quantum learning means finding the optimal storing board and the optimal retrieving channel.

An alternative to coherent retrieval is to estimate $U$, to store the
outcome in a classical memory, and to perform the estimated unitary on
the new input state.  This incoherent estimation-based strategy has the double
advantage of avoiding the expensive use of a quantum memory (which
nowadays cannot store information for more than few milliseconds),---and
of allowing one to reproduce $U$ an unlimited number of times with
constant quality.  However,   estimation-based strategies  are typically suboptimal for the similar task of quantum cloning \cite{clonrev},  and, by analogy, one  would expect a coherent retrieval to achieve better performances.
Surprisingly, we find that  whenever the unknown unitary is randomly drawn from a group  the incoherent strategies already achieve
the ultimate performances for quantum learning.  In particular, we show that  the performance of the optimal retrieving channel is equal to that of optimal estimation.  For example, for a completely unknown qubit unitary the optimal
fidelity behaves as $F=1-O(N^{-2})$ asymptotically for large $N$. Our
result can be also extended to solve the problem of \emph{optimal
  inversion} of the unknown $U$, in which the user is asked to perform
$U^\dag$. In this case, we provide the optimal approximate
realignement of reference frames for the quantum communication
scenario considered by Ref.  \cite{qcommrefframe}, reaching the above
asymptotic fidelity without ancilla.  The paper is structured as follows:  in Sec. \ref{sec:combs} we introduce the notation and the theoretical framework used to solve the problem of optimal learning.   
 The optimization is then presented in Sec. \ref{sec:optimization}, by first addressing the case of a single output copy (Subsect. \ref{M=1}), and subsequently showing how to generalize the argument to the case of $M>1$ output copies (subsection \ref{M>1}).   In Sec. \ref{sec:inv} we discuss the problem of the optimal inversion of an unknown quantum dynamics, which can be regarded as a small variation of our learning problem.   Sec. \ref{sec:conclusions} concludes the article with a summary of the main results. 

\section{Notation and theoretical framework}\label{sec:combs}

To derive the optimal learning we use the method of {\em quantum combs} \cite{architecture}, briefly summarized here. For more details and for an extensive presentation of the method we refer to Ref. \cite{comblong}.

Let ${\rm Lin} (\hilb H)$ denote the space of linear operators acting on the Hilbert space $\hilb H$, and $\rm Lin (\hilb H,\hilb K)$ be the space of linear operators from $\hilb H$ to $\hilb K$.  In the following we will use the
one-to-one correspondence between bipartite vectors  $|A\kk \in \hilb K \otimes \hilb H$ and linear operators
$A \in {\rm Lin} (\hilb H,\hilb K)$ given by 
\begin{equation}\label{doubleket} 
|A\kk = \sum_{m =1}^{{\rm dim} (\hilb K)} \sum_{n=1}^{{\rm dim} (\hilb H)}\<m |A|n\> ~ |m\> |n\>,
\end{equation} 
where $\{|m\> \}_{m=1}^{{\rm dim} (\hilb K)}$ and $\{|n\> \}_{n=1}^{{\rm dim} (\hilb H)}$ are two fixed orthonormal bases for $\hilb K$ and $\hilb H$, respectively.  

 If $A$ and $B$ are two commuting operators in  $ {\rm Lin} (\hilb H)$ it is simple to derive from Eq. (\ref{doubleket}) the equality
\begin{equation}\label{mirrorket}
(A \otimes I_{\hilb H}) |B\kk = (I_{\hilb H}\otimes A^T) |B\kk~, 
\end{equation} 
where $I_{\hilb H}$ is the identity operator on $\hilb H$ and $A^T$ denotes the transpose of $A$ with respect to the orthonormal basis $\{|n\>\}$.

A quantum channel  $\map C$ from ${\rm Lin} (\hilb H)$ to ${\rm Lin} (\hilb K)$ is a completely positive trace-preserving map, and  is conveniently described by its
Choi-Jamio\l kowski operator, namely by the positive operator $C\in {\rm Lin} (\sK\otimes\sH) $ defined by 
\begin{equation}\label{choi} 
C = (\map C \otimes \map I_{\hilb H}) (|I_{\hilb H}\kk \bb I_{\hilb H}| ) ,
\end{equation}
where $\map I_{\hilb H} $ is the identity map on ${\rm Lin } (\hilb H)$, and, according to Eq. (\ref{doubleket}), $|I_{\hilb H}  \kk$ is the maximally entangled vector $|I_{\hilb H }\kk = \sum_{n=1}^{{\rm dim} (\hilb H)} |n\>|n\>\in\sH^{\otimes2}$. 

The composition of two channels is represented in terms of their Choi-Jamio\l kowski operators by the \emph{link product} \cite{architecture,comblong}. Precisely, if $\map D$ is a channel from $\hilb K$ to
$\hilb L$, the Choi operator of the channel $\map D \circ\map C$
resulting from the composition of  $\map C$ and $\map D$ is given by the product 
\begin{equation}\label{link} 
  D *
  C = \Tr_{{\hilb K}} [(D \otimes I_{\hilb H}) (I_{\hilb L} \otimes C^{T_{\hilb K}}) ]~,
\end{equation} 
with $\Tr_{\hilb K}$ denoting partial transpose on $\hilb K$.   Viewing states as a special kind of channels with one-dimensional
input space,  Eq. (\ref{link}) yields $\map C (\rho) = C
* \rho = \Tr_{\hilb H}[ C (I_{\hilb K} \otimes \rho^T)]$.  A channel $\map C$ from
$\hilb H $ to $\hilb K$ is trace preserving if and only if it
satisfies the normalization condition 
\begin{equation}\label{channelnorm}
 I_{\hilb K} * C \equiv \Tr_{\hilb K}[C] =
I_{\hilb H}.\end{equation}

For two channels with multipartite input and output, one can decide to connect  only some particular output of the first channel to some input of the second one: for example, if $\map C$ is a channel from ${\rm Lin} (\hilb H \otimes \hilb A)$ to   ${\rm Lin} (\hilb K \otimes \hilb B)$ and $
\map D$ is a channel from ${\rm Lin} (\hilb A' \otimes \hilb K)$ to ${\rm Lin} (\hilb B' \otimes \hilb L)$
 we can connect the wires with the same label $\hilb K$, thus obtaining the new channel  $(\map D 
 \otimes \map I_{\hilb B}) (\map I_{\hilb A'} \otimes \map C)$,  which is a channel from ${{\rm Lin} (\hilb A' \otimes \hilb H \otimes \hilb A)}$ to  ${{\rm Lin} (\hilb B' \otimes \hilb L \otimes \hilb B)}$. 
 Accordingly, the connections of quantum channels in a network will be encoded in the labels assigned to the Hilbert spaces: whenever two spaces have the same label,  two channels acting on these spaces will be connected, and their Choi-Jamio\l kowski operators will be contracted with the link product as in Eq. (\ref{link}).  
 
{\bf Remark (reordering of Hilbert spaces and commutativity of the link product).}  
Encoding the connections in the labeling of the Hilbert spaces turns out to be very convenient in the treatment of multipartite quantum networks, because some formulae take a much simpler form if we suitably rearrange the ordering of the Hilbert spaces in the tensor product. For example, it may be convenient to rewrite the tensor product $\bigotimes_{i=1}^{2N+1} \hilb H_i$ putting all spaces with even labels on the left and all spaces with odd labels on the right. This reordering  can be done safely as long as different Hilbert spaces have different labels.   Note that the link product of two Choi-Jamio\l kowski operators is commutative up to this reordering of Hilbert spaces: for example, given two operators $C \in {\rm Lin}  (\hilb K \otimes \hilb H) $    and   $D \in {\rm Lin}  (\hilb L \otimes \hilb K) $ with $\hilb H \simeq \hilb K \simeq \hilb L$,   we have $D*C = {\rm SWAP} ~ (C*D )~{\rm SWAP}$, where ${\rm SWAP}$ is the operator that exchanges the Hilbert spaces $\hilb L $ and $\hilb H$  in the tensor product  $\hilb L \otimes \hilb H$.    The reader should not be confused by fact that the link product is commutative (up to reordering of the Hilbert spaces) whereas the composition of channels is not  ($\map C \circ \map D $ is in  general different from $\map D \circ \map C$). The fact that the output of $\map C$ is connected with the input of $\map D$ (and not the other way round)  is encoded in the fact that the  output space of $\map C$ has the same label  of the input space of $\map D$ (here they are both labeled as $\hilb K$).  In order to express the different composition of channels corresponding to $\map C \circ \map D$  we would have had to choose a different labeling, in which the output of $\map D$ is identified with the input of $\map C$.      
 
 \medskip
A quantum circuit board is the quantum network resulting from a sequence of multipartite channels where some input of a channel is connected to some output of the previous one, as we just illustrated.   A \emph{quantum comb} is the Choi-Jamio\l kowski operator associated to a quantum circuit board, and is obtained as the link product of all component channels. The fact that the the circuit board represents a sequence of (trace-preserving) channels is expressed by a set of linear equations \cite{architecture,comblong}, and, therefore,  optimizing a quantum circuit board is equivalent to optimizing a positive operator subject to these linear constraints.  The constraints will be given explicitly for the case of learning in the next section.

\section{ Optimization of learning}\label{sec:optimization}  
In this section we show that the optimal quantum learning of  an unknown unitary  randomly drawn from a group has a very simple and general structure: \emph{(i)}  in order to store the unitary it is enough to apply  the available examples in parallel on a suitable entangled state, \emph{(ii)} the optimal state for storage has the same form of an optimal state for estimation of the unknown unitary, and  \emph{(iii)}  the optimal retrieval can be achieved via estimation of the unknown unitary, namely  by measuring the quantum memory, producing an estimate for the unknown unitary, and finally, applying the estimate $M$ times.

\subsection{The  $M =1$ case}\label{M=1}
\begin{figure}[t]
\includegraphics[width= \columnwidth]{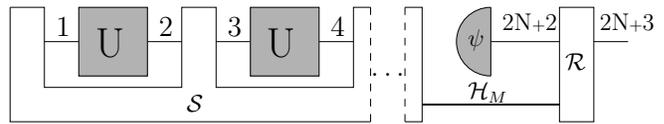}
\caption{The learning process is described by a quantum comb (in white)
 representing the storing board, in which the $N$ uses of a unitary
 $U$ are plugged, along with the state $|\psi\>$ (in gray).  The wires
  represent the input-output Hilbert spaces. The output of the first comb is 
  stored in a quantum memory, later used by the retrieving channel $\map R$. 
  \label{f:comb}}
\end{figure}

We tackle the optimization of learning starting from the case where a single output copy is required.
Referring to Fig.  \ref{f:comb}, we label the Hilbert spaces of
quantum systems according to the following sequence:
$(\hilb{H}_{2n+1})_{ n=0}^{N-1}$ are the inputs for the $N$ examples
of $U$, and $(\hilb{H}_{2n+2})_{n=0}^{N-1}$ are the corresponding
outputs. We denote by $\hilb H_i = \bigotimes_{n=0}^{N-1} \hilb H_{2n
  +1}$ ( $\hilb H_o = \bigotimes_{n=0}^{N-1} \hilb H_{2n+2}$) the
Hilbert spaces of all inputs (outputs) of the $N$ examples. The input
state $|\psi\>$ belongs to $\hilb{H}_{2N+2}$, and the output state
finally produced belongs to $\hilb H_{2N+3}$.  All spaces $\hilb H_n $
considered here are $d-$dimensional, except the spaces $\hilb H_0$ and
$\hilb H_{2 N +1}$ which are one-dimensional and are introduced just
for notational convenience.  The comb of the whole learning process is
an operator $L\geq 0$ on the tensor of all Hilbert spaces and satisfies
the normalization condition \cite{architecture, comblong}:
\begin{equation}\label{recnorm}
\Tr_{2k+1} [ L^{(k)}] = I_{2k} \otimes L^{(k-1)} \qquad k=0,1, \dots, N+1~
\end{equation} 
where $L^{(N+1)}=L$, $L^{(-1)} =1$, and $L^{(k)}$ is a positive
operator on the spaces $(\hilb H_n)_{n=0}^{2k+1}$. When the $N$
examples are connected with the learning board, the user obtains a
channel $\map C_U$ with Choi operator given by 
\begin{equation}
\begin{split}
C_U &= L * |U\kk \bb
U|^{\otimes N} \\
&= \Tr_{i,o} \left[ L \left( I_{2N+3} \otimes I_{2N+2}
    \otimes (|U\kk \bb U|^{ \otimes N})^T \right) \right],
 \end{split}
  \end{equation} 
as it follows from the definition of link product in Eq. (\ref{link}).

As the figure of merit we maximize the fidelity of the output state $\map{C}_U(|\psi\>\<\psi|)$ with
the target state $U\ket{\psi}\bra{\psi} U^\dag$, uniformly averaged over all input pure states
$\ket{\psi}$ and all unknown unitaries $U$ in the group $G$.  Apart from irrelevant constants, such
optimization coincides with the maximization of the channel fidelity between $\map C_U$ and the
target unitary ({\em i.e.} the fidelity between the Choi-Jamio\l kowski states $C_U/d$ and $|U\kk \bb U| /d$)
averaged over $U$:
\begin{equation}\label{figmer}
\begin{split}
 F&=    \frac 1 {d^2} ~
 \int_{G}   \Tr  \left\{ L \left [   |U\kk \bb U| 
 \otimes (|U\kk \bb U|^{ \otimes N})^T \right] \right\} \d U\\ 
&=    \frac 1 {d^2} ~
\int_{G} \Bra{U}\Bra{U^*}^{\otimes N} L \Ket{U^*}^{\otimes N}\Ket{U} \, \d U~,
\end{split}
\end{equation}
$U^* $ being the complex conjugate of $U$ in the computational basis,
and $\d U$ denoting the normalized Haar measure.  From the expression
of $F$ it is easy to prove that there is no loss of generality in
requiring the commutation
\begin{equation} 
 [L, U_{2N+3} \otimes V^*_{2N+2} \otimes 
 (U^*  \otimes V)^{\otimes N} ] = 0  \qquad \forall U, V \in G~. \label{covL}
\end{equation}

Moreover, using Eq.  (\ref{recnorm}) for $k =N+1$ we obtain $\Tr_{\hilb H_{2N+3}}  [ L] = I_{2N+2} \otimes L^{(N)}$, where $L^{(N)}$ is a positive operator acting on $\bigotimes_{n=0}^{2N+1} \hilb H_n$ (recall that, however, $\hilb H_0$ and $\hilb H_{2N+1}$ are one-dimensional). Reordering the Hilbert spaces in the tensor product by putting all input spaces of the examples on the right and all output spaces on the left and using Eq. (\ref{covL})  we then get
\begin{equation} 
 [L^{(N)}, U^{*\otimes N}_o \otimes V^{\otimes N}_i] = 0 \qquad \forall U, V\in G. 
 \label{covred}
\end{equation}
Here the subscripts $i,o$ recall that $U^{\otimes N}$ acts on the tensor product of all output spaces $\hilb H_o = \bigotimes_{n=0}^{N-1}  \hilb H_{2n+1}$, while  $V^{\otimes N}$ acts on the tensor product of all input spaces $\hilb H_i = \bigotimes_{n=0}^{N-1}  \hilb H_{2n+1}$.  
This leads to the following
\begin{lemma}[Optimality of parallel storage] The optimal storage of
  $U$ can be achieved by applying $U_o^{\otimes N} \otimes I_i^{\otimes
    N}$ on a suitable input state $|\varphi \> \in \hilb H_{o} \otimes
  \hilb H_{i}$.
\end{lemma}

\emph{Proof.} According to Fig.   \ref{f:comb}, the learning board $\map L$ results from the connection of the
storing board $\map S$ with the retrieving channel $\map R$. In terms of the corresponding Choi-Jamio\l kowski operators $L,  S, R$, respectively,  one has $L
= R * S $. Denoting by $\hilb H_M$ the Hilbert space of the quantum
memory in   Fig. \ref{f:comb},  we have that $\map R$ is a channel from $\left(\hilb H_{2N+2} \otimes \hilb
  H_M\right)$ to $\hilb H_{2N+3}$, and satisfies the normalization
condition $ I_{2N+3} * R = I_{2N+2} \otimes I_M$. Using this fact, one
gets $\Tr_{2N+3} [L ] \equiv I_{2N+3 } * L = (I_{2N+3} * R) * S =
(I_{2N+2} \otimes I_M) * S= I_{2N+2} \otimes \Tr_M[ S]$, which
compared with Eq.  (\ref{recnorm}) for $k=N+1$ implies $\Tr_M[S] =
L^{(N)}$. Now, without loss of generality we take the storing board
$\map S$ to be a sequence of isometries \cite{architecture,comblong}, which
implies that $S$ is rank one: $S = |\Phi\kk \bb \Phi|$. With this
choice, the state $S /d^N$ is a purification of $L^{(N)}/d^N$.  Again,
one can choose w.l.o.g.  $S/d^N$ to be a state on $(\hilb H_o \otimes
\hilb H_i) \otimes (\hilb H_o' \otimes \hilb H_i')$, with $\hilb H_o'
\simeq \hilb H_o $ and $\hilb H_i' \simeq \hilb H_i$ and assume $|\Phi\kk
= | L^{(N) \frac 1 2} \kk $. Taking $V =I$ in Eq.  (\ref{covred}) and
using Eq.  (\ref{mirrorket}) we get $ \left(U_o^{\otimes N } \otimes
  I_{i, o',i'}\right) |\Phi\kk =\left(I_{o,i} \otimes
  {U_{o'}^T}^{\otimes N } \otimes I_{i'}\right )|\Phi\kk $. When the
examples of $U$ are connected to the storing board, the output is the
state $\rho_U = S * |U\kk \bb U|_{o,i}^{\otimes N}$.  Using the above
relation we find that $\rho_U$ is the projector on the state $
|\varphi_U\>= (U_{o'}^{\otimes N} \otimes I_{i'}) |\varphi \> $, where
$|\varphi \> = \bb I^{\otimes N} |_{o,i} |\Phi\kk \in \hilb H_{o'}
\otimes \hilb H_{i'} \simeq \hilb H_o \otimes \hilb H_i$.  This proves that every storing board gives the
same output that would be obtained with a parallel scheme. In other words, every storing board can be simulated applying $(U_o^{\otimes N} \otimes I_i^{\otimes N})$ to a suitable input state $|\varphi\>\in \hilb H_o \otimes \hilb H_i$. \qed
 
Optimizing learning is then reduced to finding the optimal input state
$|\varphi \>$ and the optimal retrieving channel $\map R$.  The
fidelity can be computed substituting $L = R * S$ in Eq.
(\ref{figmer}) and using the relation $\bb U| \bb U^* |^{\otimes N}
(R*S) |U\kk | U^* \kk^{\otimes N} = \bb U| R |U\kk * \bb U^*|^{\otimes
  N} S |U^*\kk^{\otimes N} = \bb U| R|U\kk * |\varphi_U\>\< \varphi_U|$, which gives
\begin{align}\label{figmer2}
  F = \frac 1 {d^2} \int_G \Bra{U} \< \varphi_U^* | R \Ket{U}
  |\varphi_U^* \> ~\d U.
\end{align}

\begin{lemma}[Optimal states for storage]\label{lemma:optstates}
  The optimal input state for storage can be taken of the form
\begin{equation}\label{inpst}
  |\varphi \> = \bigoplus_{j\in {\rm Irr} (U^{\otimes N})} \sqrt{\frac{p_j}{d_j}}  |I_j\kk  \in \widetilde{\hilb H}
  ~,\end{equation} 
where $p_j$ are probabilities, the index $j$ runs over the set ${\rm Irr} (U^{\otimes N})$
 of all irreducible representations  $\{U_j\}$ contained in the decomposition of $\{U^{\otimes N}\}$, and $\widetilde{\hilb H} = \bigoplus_{j\in {\rm Irr} (U^{\otimes N}) } (\hilb H_j \otimes \hilb
H_j)$ is a subspace of $\hilb H_o \otimes \hilb H_i$ carrying the
representation $\widetilde{U} = \bigoplus_{j\in {\rm Irr (U^{\otimes N})}}( U_j \otimes I_{j})$,  $I_j$ being the identity in $\hilb H_j$.
\end{lemma}
\emph{Proof.}  Using Eqs. (\ref{mirrorket}) and (\ref{covred}) it is possible
to show that the marginal state $\rho = \Tr_i[|\varphi \>\< \varphi |]$
is invariant under $U^{\otimes N}$. Decomposing $U^{\otimes N}$ into
irreducible representations (irreps) we have $U^{\otimes N } =
\bigoplus_j (U_j \otimes I_{m_j})$, where $I_{m_j}$ is the identity on
an $m_j$-dimensional multiplicity space $\mathbb C^{m_j}$.  Therefore,
$\rho$ must have the form $ \rho= \bigoplus_j p_j (I_j/ d_j \otimes
\rho_{j})$, where $\rho_{j}$ is an arbitrary state on the multiplicity
space $\mathbb C^{m_j}$.  Since $|\varphi \>$ is a purification of
$\rho$, with a suitable choice of basis we have $|\varphi \>=
|\rho^{\frac 1 2} \kk = \bigoplus_j \sqrt{p_j / d_j}~ |I_j \kk
|\rho_{j}^{\frac 1 2} \kk $, which after storage becomes
$|\varphi_U\>= \bigoplus_j \sqrt{p_j/d_j}|U_j \kk |\rho_{j}^{\frac 1
  2} \kk$.  Hence, for every $U$ the state
$|\varphi_U\>$ belongs to the subspace $\widetilde{\hilb{H}} =
\bigoplus_{j} ( \hilb H_j^{\otimes 2} \otimes |\rho^{\frac 1
  2}_{j}\kk) \simeq \bigoplus_j \hilb H_j^{\otimes 2}$.  \qed

We can then restrict our attention to the subspace $\widetilde {\hilb
  H}$, and consider retrieving channels $\map R$ from $(\hilb H_{2N
  +2} \otimes \widetilde{\hilb{H}})$ to $\hilb H_{2N+3}$. The
normalization of the Choi operator is then 
\begin{equation}\label{normRetrieving}
\Tr_{2N+3} [R] = I_{2N+2}
\otimes I_{\widetilde {\hilb H}}~.
\end{equation} Combining the expression of the
fidelity (\ref{figmer}) with that of the input state (\ref{inpst}), it
is easy to see that one can always use a covariant retrieving channel, satisfying 
\begin{align}\label{cov}
  \left[R, U_{2N+3} \otimes V^*_{2N+2} \otimes \widetilde U^* \widetilde
    V' \right] = 0 \qquad \forall U, V \in G
\end{align}
where $\widetilde V' = \bigoplus_j (I_j \otimes V_j)$ acts on
$\widetilde {\hilb H}$.  
We now exploit the decompositions $U \otimes U_j^* = \bigoplus_{K\in {\rm Irr} (U\otimes U_j^*)}
\left(U_K \otimes I_{m^{(j)}_K}\right) $ and $V^* \otimes V_j =
\bigoplus_{L\in {\rm Irr (V^*\otimes V_j)}} \left(V^*_L \otimes I_{m_L^{(j)}}\right)$, which yield
\begin{equation}\label{decompUVUV}
  U_{2N+3} \otimes V^*_{2N+2} \otimes \widetilde{U}^* \widetilde{V} = \bigoplus_{K,L} \left( U_K \otimes V^*_L \otimes I_{m_{KL}}\right)~.
\end{equation} 
Here $I_{m_{KL}}$ is given by $I_{m_{KL}} = \bigoplus_{j \in
  \set{P}_{KL}}\left( I_{m_K^{(j)}} \otimes I_{m^{(j)}_L}\right)$,
where $\set P_{KL}$ is the set of values of $j$ such that the irrep
$U_K \otimes V^*_L$ is contained in the decomposition of $U \otimes
V^* \otimes U^*_j \otimes V_j$.  Relations (\ref{cov}) and
(\ref{decompUVUV}) then imply
\begin{equation}\label{BlockR} R = \bigoplus_{K,L} (I_K
  \otimes I_L \otimes R_{KL})~,
\end{equation} 
where $R_{KL}$ is a positive operator on the multiplicity space
$\mathbb C^{m_{JK}} = \bigoplus_{j \in \set P_{KL}} \left( \mathbb
  C^{m_K^{(j)}} \otimes \mathbb C^{m_{L}^{(j)}} \right)$. Moreover,
using the equality $I \otimes I_j =\bigoplus_{K} (I_{K} \otimes
I_{m_K^{(j)}})$ we obtain
\begin{equation}\label{phi-id}
\begin{split}
   \Ket{I} |\varphi^*\>&= \bigoplus_{j} \sqrt{\frac{p_j} {d_j}} \Ket{I} \Ket{I_j} \\
   & = \bigoplus_{j} \bigoplus_{K \in{ \rm  Irr}  (U\otimes U^*_j)}   \sqrt{\frac{p_j} {d_j}}  \Ket{I_K} \Ket{I_{m_K^{(j)}}} \\
  & = \bigoplus_{K} \bigoplus_{j \in \set P_{KK}}   \sqrt{\frac{p_j} {d_j}}  \Ket{I_K} \Ket{I_{m_K^{(j)}}} \\
   & = \bigoplus_{K}  |I_K \kk |\alpha_K \>~,
\end{split}
\end{equation}
where $|I_K\kk \in \hilb H_K^{\otimes 2}$ and $|\alpha_K\> \in \mathbb C^{m_{KK}} $ is given by
\begin{equation}
|\alpha_K\> =\bigoplus_{j\in \set P_{KK}}\sqrt{p_j/d_j}\ |I_{m_K^{(j)}} \kk.
\end{equation}
Exploiting Eqs. (\ref{BlockR}) and (\ref{phi-id}),  the fidelity
(\ref{figmer2}) can be rewritten as
\begin{equation}\label{figmer3}
F= \sum_K  \frac {d_K}{d^2}~ \<\alpha_K | R_{KK} |\alpha_K \>~.
\end{equation}

\begin{theorem}[Optimal retrieving strategy]\label{theo:OptRetr} The optimal retrieving of
  $U$ from the memory state $|\varphi_U \>$ is
  achieved by measuring the ancilla with the optimal POVM $P_{\hat U} =
  |\eta_{\hat U} \>\< \eta_{\hat U}|$ given by $|\eta_{\hat U}\> =
  \bigoplus_j \sqrt {d_j} |\hat U_j\kk$, and, conditionally on outcome $\hat U$,
  by performing the unitary $\hat U$ on the new input system.
\end{theorem}

\emph{Proof.} Let us denote by $P^{(j)}_{KL}$ the projector on the tensor
product $\mathbb C^{m^{(j)}_K} \otimes \mathbb C^{m_L^{(j)}}$, and by
$R^{(j)}_{KL} = P^{(j)}_{KL} R_{KL} P^{(j)}_{KL}$ the corresponding
diagonal block of $R_{KL}$. Using Schur's lemmas and Eq.
(\ref{BlockR}) we obtain 
\begin{equation}
\Tr_{2N+3} [ R] =\sum_{ K,L} \sum_{j \in
 \set P_{KL}} \left(\frac{d_K}{d_j} I_j \otimes I_L \otimes
  \Tr_{m_K^{(j)}} [ R^{(j)}_{KL} ]\right).
  \end{equation}
  Equation (\ref{normRetrieving}) then becomes 
$I_{m^{(j)}_L}=\sum_{K|\set P_{KL}\ni j} \frac{d_K}{d_j}~ \Tr_{m_K^{(j)}} [
R^{(j)}_{KL} ]$ for all $L,j$, which for $K=L$ implies the bound
\begin{equation}\label{normgroup}
\Tr[R_{KK}^{(j)}] \le \frac{d_j m_K^{(j)}}{d_K}~.
\end{equation}   For the fidelity
(\ref{figmer3}) we then have the bound
\begin{align}
  F & = \sum_{K}  \frac{d_K} {d^2} \sum_{j, j' \in \set P_{KK}} \sqrt{\frac{p_j p_{j'}}{d_{j} d_{j'}}}  \bb I_{m^{(j)}_K} |R_{KK} |I_{m^{(j')}_K}\kk\\
   & \le \sum_K \frac{d_K} {d^2} \left( \sum_{j \in \set P_{KK}} \sqrt{
      \frac{ p_j \bb I_{m_K^{(j)}}| R^{(j)}_{KK}
        |I_{m_K^{(j)}}\kk }{d_{j}}}\right)^2\\
 \label{FidBound} & \le \sum_K \frac { \left(\sum_{j \in \set P_{KK}} m_K^{(j)} \sqrt
  {p_j}\right)^2}{d^2} =F_\mathrm{est}~,
\end{align}
having used the positivity of $R_{KK}$ for the first bound and Eq.
(\ref{normgroup}) for the second.  Regarding the last equality, it can
be proved as follows.  First, the Choi operator of the
estimation-based strategy is $R_\mathrm{est} = \int_G |\hat U\kk\bb
\hat U| \otimes |\eta^*_{\hat U} \>\< \eta^*_{\hat U} | \d \hat U$.
Using Eq.  (\ref{phi-id}) with $|\varphi^*\>$ replaced by
$|\eta_{I}^*\>$ and performing the integral we obtain $R_\mathrm{est}
= \bigoplus_K (I_{K}^{\otimes 2} \otimes \widetilde{R}_{KK})/d_K $,
where $\widetilde R_{KK} = |\beta_K\>\<\beta_K|$, $|\beta_K\> =
\bigoplus_{j \in \set P_{KK}} \sqrt{d_j} |I_{m_K^{(j)}}\kk$. Eq.
(\ref{figmer3}) then gives 
\begin{equation}
\begin{split}
F_\mathrm{est} &= \sum_{K} ~
\frac{|\<\alpha_K|\beta_K\>|^2}{d^2} \\
&= \sum_K \frac { \left(\sum_{j \in \set P_{KK}} m_K^{(j)} \sqrt
  {p_j}\right)^2}{d^2}.
\end{split}
\end{equation} \qed

The above theorem shows that the optimal state for storing $U$ is
identical to the optimal state for estimating it \cite{EntEst}, and, moreover, that the fidelity of unitary learning with $M=1$ is precisely the fidelity of unitary estimation.  Having reduced learning to estimation, we can then exploit the expressions for the optimal states and fidelities that are known in most relevant cases. 
  For example, when $U$ is an unknown qubit unitary in ${\rm SU}(2)$, learning becomes equivalent to optimal
estimation of an unknown rotation in the Bloch sphere \cite{refframe}.
For large number of copies, the optimal input state is given by
$|\varphi\>\approx \sqrt{4/N}~\sum_{j= j_{\min}}^{N/2} \frac{\sin
  (2\pi j/N)}{\sqrt{2j+1}} ~ |I_{j}\kk$, with $j_{\min} = 0 (1/2)$ for
$N$ even (odd), and the fidelity is $F\approx 1-\pi^2/4N^2$.
Remarkably, this asymptotic scaling can be achieved without using
entanglement between the set of $N$ qubits that are rotated and an
auxiliary set of $N$ rotationally invariant qubits: the optimal
storing is achieved just by applying $U^{\otimes N}$ on the optimal
$N$-qubit state \cite{refframe}. Another example is that of an unknown
phase-shift $U= \exp[i \theta \sigma_z]$. In this case, for large
number of copies the optimal input state is $|\varphi\> =
\sqrt{2/(N+1)}\sum_{m=-N/2}^{N/2} \sin[\pi(m+1/2)/(N+1)] |m\> $ and
the fidelity is $F\approx 1-2\pi^2/(N+1)^2$ \cite{OptClocks}. Again,
the optimal state can be prepared using only $N$ qubits.

\subsection{Generalization to the $M>1$ case}\label{M>1}

Our result can be extended to the case where the user must reproduce
$M>1$ copies of the unknown unitary $U$.   In this case, there are two different notions of optimality induced by two different figures of merit, namely the single-copy  and the global fidelity. In the following we will examine both cases.

\subsubsection{Optimal learning according to the single-copy fidelity}
Let $\map C_U$ be the
$M$-partite channel obtained by the user, and $\map C_{U,\Sigma}^{(1)}$ be the
local channel $\map C_{U,\Sigma}^{(1)} (\rho) = \Tr_{\bar 1}[\map C_U (\rho
\otimes \Sigma)]$, where $\rho$ is the state of the first system, $\Sigma$ is the state of the remaining $M-1$ systems, and $\Tr_{\bar 1}$ denotes the trace over all
systems except the first. The local channel $\map C_{U,\Sigma}^{(1)}$ describes
the evolution of the first input of $\map C_U$ when the remaining
$(M-1)$ inputs are prepared in the state $\Sigma$.  Of course, the
fidelity between $\map C_{U,\Sigma}^{(1)}$ and the unitary $U$ cannot be larger
than the optimal fidelity $F_\mathrm{est}$ of Eq.  (\ref{FidBound}),
and the same holds for any local channel $\map C^{(i)}_{U,\Sigma}$, in which
all but the $i$-th input system are discarded.  Therefore, the
measure-and-prepare strategy presented in Theorem \ref{theo:OptRetr} is optimal also for the maximization of
the single-copy fidelity of all local channels, and such fidelity does
not decrease with increasing $M$.  
\subsubsection{Optimal learning according to the global fidelity} 
The results of subsection \ref{M=1}   can be
extended to the maximization of the global fidelity between $\map C_U$
and $U^{\otimes M}$, just by replacing $U$ with $U^{\otimes M}$ in all
derivations.  Indeed, the role of the target unitary $U$ in our derivations is completely generic: we never used the fact that the unitary emulated by the machine was equal to the unitaries provided in the examples.   
Therefore, following the same proofs of subsection \ref{M=1} it is immediate to see that also for the case of $M>1$ copies with global fidelity the optimal strategy for storing consists in the parallel application of the examples on an input state of the form of Lemma \ref{lemma:optstates} and that the optimal strategy for retrieving consists in measuring
the optimal POVM $P_{\hat U} $ and in performing $\hat U^{\otimes M}$
conditionally on outcome $\hat U$.   Therefore, also in this case optimal learning is equivalent to optimal estimation: precisely, the optimal learning is achieved by the estimation strategy that maximizes the expectation value of the goal function  $f_M (U, \hat U) = (|\Tr[U^\dag \hat U]|/d)^{2M}$, given by
$\<f_M\>  = \int \d U \int \d \hat U   ~f_M (U, \hat U) ~  \< \varphi_U |  P_{\hat U}  |\varphi_U\>$.  Note that in this case the coefficients $\{p_j\}$ in the optimal input state of Lemma \ref{lemma:optstates}) generally depend on $M$. 

\medskip
 {\bf Remark (generalization to nonidentical group representations).}  Since we never used the fact that the $N$ examples are identical, all the results of Subsect. \ref{M=1}   hold even when the input (output) uses are not identical copies
$U^{\otimes N}$ ($U^{\otimes M}$), but generally $N$ ($M$) different
unitaries, each of them belonging to a different representation of the
group $G$. For example, if $G={\rm SO}(3)$  the $N$ examples may correspond to rotations (of the same angle and around the same axis) of $N$ quantum particles with different  angular momenta.   Of course, the same remark also holds when the $M$ output copies. 

\section{Optimal inversion of an unknown unitary evolution}\label{sec:inv}

We now extend our results to the \emph{optimal inversion} of an unknown unitary $U$: in this case the goal is not to produce $M$ copies of $U$, but, instead $M$ copies of its inverse $U^\dag$.  For
this task the fidelity of the learning board is $F' = 1/d^2 \int_G \bb U^{\dag} |^{\otimes M} \bb
U^*|^{\otimes N} ~ L'~|U^\dag\kk^{\otimes M} |U^*\kk^{\otimes N} \d U$, as obtained by substituting
$U$ with $U^{\dag \otimes M}$ in the target of Eq.  (\ref{figmer}).  From this expression it is easy
to see that one can always assume $[L', V^{\otimes M} \otimes U^{*\otimes M}\otimes U_o^{*\otimes N}
\otimes V_i^{\otimes N} ]=0$.  Therefore, the optimal inversion is obtained from our derivations by
simply substituting $U_{2N+3} \to V^{\otimes M}$ and $V_{2N+2} \to U^{\otimes M}$.  Accordingly, the
optimal inversion is achieved by measuring the optimal POVM $P_{\hat U}$ on the optimal state
$|\varphi_U\>$ and by performing $\hat U^{\dag \otimes M}$ conditionally on outcome $\hat U$. This
provides the optimal approximate realignement of reference frames in the quantum communication
scenario recently considered in Ref.  \cite{qcommrefframe}, proving the optimality of the ``measure-and-rotate" strategy conjectured therein. In that scenario, the state $|\varphi\> \in
\widetilde{\hilb H}$ serves as a token of Alice's reference frame, and is sent to Bob along with
a quantum message $|\psi\> \in \hilb H^{\otimes M}$.  Due to the mismatch of reference frames, Bob
receives the decohered state $\sigma_\psi = \int_G |\varphi_U\>\<\varphi_U| \otimes U
|\psi\>\<\psi|U^\dag \d U$, from which he tries to retrieve the message $|\psi\>$ with maximum
fidelity $f = \int \d \psi~ \<\psi| \map R' (\sigma_\psi)|\psi\> \d \psi$, where $\map R'$ is the
retrieving channel and $\d \psi$ denotes the uniform probability measure over pure states.  The
maximization of $f$ is equivalent to the maximization of the channel fidelity $F' = \int_G \bb
U^\dag| \< \varphi_U^*| R' |U^\dag\kk |\varphi_U^*\> \d U$, which is the figure of merit for optimal
inversion.  It is worth stressing that the state $|\varphi\>$ that maximizes the
fidelity is not the state $|\varphi_\mathrm{lik}\> = \bigoplus_j \sqrt{d_j/L} |I_j\kk$, $L=\sum_j
d_j^2$ that maximizes the likelihood \cite{covlik}. For $M =1$ and $G = SU (2), U(1)$ the state
$|\varphi\>$ gives an average fidelity that approaches 1 as $1/N^2$, while for
$|\varphi_\mathrm{lik}\>$ the scaling is $1/N$.  On the other hand, Ref. \cite{qcommrefframe} shows
that for $M=1$ $|\varphi_\mathrm{lik}\>$ allows a perfect correction of the misalignment errors with
probability of success $p = 1- 3/(N+1)$, which is not possible for $|\varphi\>$. The
determination of the best input state to maximize the probability of success, and the study of the
probability/fidelity trade-off remain open interesting problems for future research.

\section{Conclusions}\label{sec:conclusions}
In conclusion, in this paper we found the optimal storing-retrieving
of an unknown group transformation with $N$ input and $M$ output
copies, proving the optimality of the incoherent ``measure-and-rotate'' strategy, in strong contrast with the case of quantum cloning. The result has been extended to the
optimal inversion of $U$, with application to the optimal approximate alignment of reference frames for quantum communication.
An interesting development of this work is the analysis of optimal learning when the unknown unitaries do not form a group. This would be the case, for example, of the optimal learning of the unknown unitary transformation appearing in Grover's quantum search algorithm. The question whether coherent quantum strategies can lead to an improvement in these cases remains open and worth investigating.

\section{Acknowledgments}
 This work has been supported by the Italian Ministry of Education through the grant PRIN 2008, and by the EC through the projects COQUIT and
CORNER. GC is grateful to R. Spekkens for useful discussions and to the Risk and Security Study Center of IUSS Pavia for financial support in the early stage of this work.  Research at Perimeter Institute for Theoretical Physics is supported in part by the Government of Canada through NSERC and by the Province of Ontario through MRI.


\begin{thebibliography}{99}
\bibitem{qmem1} B. Julsgaard, \emph{et al.}, Nature 432, 482 (2004).
\bibitem{qmem2} C. F. Roos, \emph{et al.}, Science {\bf 304} 1478 (2004).
\bibitem{qmem3} P. Rabl, \emph{et al}, Phys.  Rev. Lett. {\bf 97} 033003  (2006).
\bibitem{cirac} G. Vidal, L. Masanes, and J. I. Cirac, Phys. Rev.
  Lett. {\bf 88}, 047905 (2002).
\bibitem{sasaki} M. Sasaki, A. Carlini, and R. Jozsa, Phys. Rev. A
  {\bf 64}, 022317 (2001).
\bibitem{sasacarl} M. Sasaki and A. Carlini, Phys. Rev. A {\bf 66},
  022303 (2002).  
\bibitem{moelmer} S. Gammelmark and K. M\o lmer, New
  J. Phys. {\bf 11} 033017 (2009).
\bibitem{clonrev}  V. Scarani \emph{et al}, Rev. Mod. Phys. {\bf 77}, 1225 (2005).
\bibitem{architecture} G. Chiribella, G. M. D'Ariano, and
  P. Perinotti, Phys. Rev. Lett. {\bf 101}, 060401 (2008). 
\bibitem{comblong}  G. Chiribella, G. M. D'Ariano, and P. Perinotti, Phys. Rev. A {\bf 80}, 022339 (2009).
\bibitem{qcommrefframe} S. D. Bartlett, T. Rudolph, R. W. Spekkens,
  and P. S. Turner, New J. Phys. {\bf 11}, 063013 (2009).
\bibitem{EntEst} G. Chiribella, G. M. D'Ariano and M. F. Sacchi,
  Phys. Rev. A 72 042338 (2005) . 
\bibitem{refframe} G. Chiribella, G. M. D'Ariano, P. Perinotti, M. F.
   Sacchi, Phys. Rev. Lett \textbf{93}, 180503 (2004).
\bibitem{OptClocks} V. Bu\u zek, R. Derka, and S. Massar, Phys. Rev.
  Lett. {\bf 82}, 2207 (1999).
\bibitem{covlik} G. Chiribella, G. M. D'Ariano, P. Perinotti, and M. F. Sacchi, Phys. Rev. A {\bf
    70}, 062105 (2004);   Int. J. Quantum Inf. 4, 453 (2006). 
\end{thebibliography}
\end{document}